\documentclass[journal]{IEEEtran}
\usepackage{multirow}
\usepackage{graphicx}
\usepackage{amsmath}
\usepackage[table,xcdraw]{xcolor}
\usepackage{url}
\ifCLASSINFOpdf
\else
\fi
\begin{document}
\title{The Value Chain of Education Metaverse}

\author{Yun-Cheng Tsai
\thanks{}}


\markboth{Journal of \LaTeX\ Class Files,~Vol.~14, No.~8, August~2015}%
{Shell \MakeLowercase{\textit{et al.}}: Bare Demo of IEEEtran.cls for IEEE Journals}



\maketitle

\begin{abstract}
Since the end of 2021, the Metaverse has been booming. Many unknown possibilities are gradually being realized, but many people only determined that they use Virtual Reality (VR), Augmented Reality (AR), and Mixed Reality (MR) in the Metaverse. It is even considered that as long as the above realities (VR, AR, MR) are used, it is equal to the Metaverse. However, this is not true, for Reality-based display tools are only one of the presentation methods of the Metaverse. If we cannot return to the three main characteristics of the Metaverse: "digital avatars," a decentralized "consensus value system," and "Immersive experience," the practice and imagination of the Metaverse will become very narrow. Since 2022, the concept of Metaverse has also been widely used in classroom teaching to integrate into teaching activities. Therefore, to prevent teachers and students from understanding the Metaverse not only in the "Using VR, AR, MR is equivalent to Metaverse" but also pay more attention to the other two characteristics of the Metaverse: "digital avatars" and a decentralized "consensus value system."
\end{abstract}


% \fi
\IEEEpeerreviewmaketitle

\section{Introduction}
\IEEEPARstart{B}{efore} the 1960s, education was knowledge-centered. Until the 1970s and 1980s, the global attention to the quality and ability of students began to become mainstream, and education began to be student-centered~\cite{ellis2014exemplars}. The fundamental reason is that traditional knowledge-centered education is increasingly difficult to meet the demand for talent in the information age, which has prompted the transformation of education into student-centered. 

The 2022 World Economic Forum (WEF) initiative "Education 4.0: learner-centered"~\cite{Education4}. According to the WEF, "Education 4.0: learner-centered" should make learner can intuitively know through appropriate additional tools. A learner-centered strategy needs to be successful auxiliary guidance. In addition to the facilitator's insight into the learner's characteristics, well-designed teaching aids and learning situations are essential items that can make learning more effective~\cite{dole2016transforming}. The current education is on the "teacher-centered" model in the age of industrial society~\cite{catalano1999transformation}. In the future, a "student-centered"  will replace the traditional teacher-centered education model. Teachers and students will be co-learners~\cite{neumann2013developing}. The instructors and guides of the school are to provide a full range of services for the growth and development of students, with the learner as the center. The traditional knowledge-based teaching model should give way to the people-centered personalized learning model. Authentic learning is not limited to one classroom or one space. Therefore, we need to reconstruct the time and length of education.

Since the end of 2021, the Metaverse has been booming~\cite{mystakidis2022metaverse}. The metaverse concept is an ongoing cyber universe that merges many virtual spaces. It is a future web iteration where users can work together, meet, play games, and socialize in these metaverse spaces~\cite{wiederhold2022ready}. If the Metaverse can extend anywhere, so can education. The future teaching model centered on classrooms, textbooks, and teachers may give way to the all-time and space-time experiential learning model based on the Metaverse. 

Since 2022, the concept of Metaverse has also been widely used in classroom teaching to integrate into teaching activities~\cite{zhai2022education}. The Metaverse also allows anyone to learn anytime, anywhere, not just in the classroom but as a collection of comprehensive human knowledge. Authentic metaverse learning will have to maximize learning, give students tasks, and enhance learning motivation.

Therefore, the future metaverse school will be fundamentally different from the current school in form, and the metaverse school is very likely to exist in the form of a "learning center"~\cite{kaput2018evidence}. In addition, domestic and foreign countries have made bold attempts at student-centered innovative schools and achieved remarkable results, which provide practical evidence for the student-centered educational theory~\cite{keengwe2009technology}.

That is to say, the static content in the original paper textbooks will be upgraded to 3D pictures, animations, audio, video, etc., to strengthen your cognition from the perspective of sight, hearing, and touch~\cite{dick2021public}. Hence, many people only determined that they use Virtual Reality (VR), Augmented Reality (AR), and Mixed Reality (MR) in the Metaverse. It even considered that as long as the above realities (VR, AR, MR) are equal to the Metaverse~\cite{buhalis2022mixed}.

However, this is not true, for Reality-based display tools are only one of the presentation methods of the Metaverse. Many unknown possibilities are gradually occurring in the Metaverse. According to Jon Radoff, the Metaverse is not 3D, 2D, or even necessarily graphical~\cite{Radoff}. If we cannot return to the three main characteristics of the Metaverse: "digital avatars," a decentralized "consensus value system," and "Immersive experience,"~\cite{chittaro2015assessing} the practice and imagination of the Metaverse will become very narrow.

Therefore, to prevent teachers and students from understanding the Metaverse, not only in the "Using VR, AR, MR is equivalent to Metaverse" but also pay more attention to the other two characteristics of the Metaverse. Based on the realization of simulation and interaction, "digital avatars" and a decentralized "consensus value system." can make the essential knowledge points in each digital textbook may also be gamified, thereby promoting students' enthusiasm for learning. Every time a student learns a new course, they can call up relevant knowledge points to remember in a time when they encounter various problems and challenges according to the gamified monster-defying upgrade process.

This paper redefines the value chain and characteristics of the "Education Metaverse." It conducts an in-depth discussion so that Education 4.0 can bring opportunities for realization due to the gradual development of the Education Metaverse, not only the integration of VR, AR, and MR into teaching but also explicitly transformed into a learner-centered learning model.

The remaining sections of this paper are as follows: Section 2 presents Metaverse's research background and value chain. Section 3 explains the value chain of the education metaverse. Section 4 is the features of the education metaverse. Section 5 concludes the article with significant findings and the possible scope of future works.

\section{Background: the Value Chain of Metaverse}
\begin{figure}[ht]
\includegraphics[width=8.8cm]{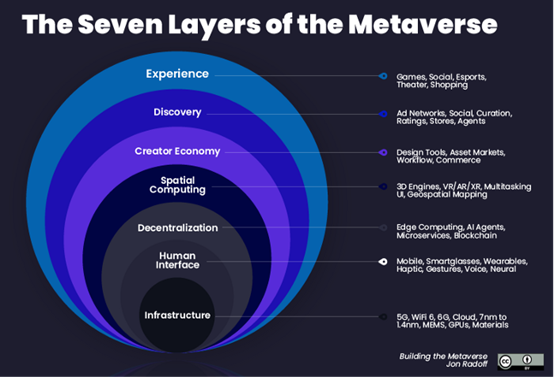}
\caption{Background: the Value Chain of Metaverse.}
\label{jon}
\end{figure}

The Metaverse Value Chain was proposed by Jon Radoff in April 2021. It's the first time an entrepreneur who has actually created Metaverse-related products presented and defined the "The Seven Layers of the Metaverse."
Figure~\ref{jon} shows the seven layers contain the following seven-point:
\begin{enumerate}
    \item Experience: At the beginning of the original text, "Many people think of the Metaverse as 3D space that will surround us. But the Metaverse is not 3D, 2D, or even necessarily graphical." The most important conclusion of this paragraph is that immersive experience is not only the experience of being immersed in the image space but also how the participants interact with the virtual world and the natural world and how to cause the proliferation and interaction of the content, and then develop into a social immersion phenomenon.
    \item Discovery: Jon Radoff refers to Discovery is to introduce new services or products to more participants. It can also be regarded as an effective promotion, which can create overall social value and thinking changes due to the participation of participants and even the shaping of culture. The digital and non-digital methods included at present can be divided into collective promotion (Inbound Marketing) and push upgrade (Outbound Marketing). The trend that comes after sharing.
    \item Creator Economy: There are content creation tools in the Metaverse to provide participants not only participants but also independent creators, sharers, and promoters, with creators as the main body and platform providers to help creators. It can focus more on serving the audience of creators and hand over financial and financial rights to creators so that they can directly raise creative funds from their audiences and provide creators with traditional methods such as advertising, placement, and sponsorship. A new opportunity.
    \item Spatial Computing: In addition to the currently recognized Metaverse using VR, AR, and MR, it echoes the original description of "Experience Layer," which is "But the Metaverse is not 3D or 2D, or even necessarily graphical." The technology that combines the world and the virtual world to achieve data integration, which can organically interact, manage and analyze data from different sources, formats, and characteristics, is the carrier of the Metaverse.
    \item Decentralization: This is one of the most prominent features of blockchain. The ideal Metaverse should be a fairer, more transparent, decentralized, and decentralized world. Jon Radoff believes that the Metaverse belongs to no one but to all at the same time.
    \item Human Interface: This is not limited to using VR, AR, MR, and other technologies in the Metaverse that everyone recognizes but refers to devices that implement VR, AR, and MR technologies, such as VR helmets. Suppose it echoes the original description of "Experience Layer," which is "But the Metaverse is not 3D or 2D, or even necessarily graphical.". In that case, mobile phones, laptops, tablets, 3D floating interactive shadow devices, smart glasses, wearable devices, etc., all belong to A human-machine interface device that realizes the Metaverse.
    \item Infrastructure: To realize the characteristics of the Metaverse mentioned above, faster and more popular network infrastructure, more extensive capacity cloud storage services, more powerful image computing equipment, high-endurance durable batteries, micro-perceptrons, and MEMS Systems, etc., all belong to the category of infrastructure.
\end{enumerate}

\section{The Value Chain of Education Metaverse}
So what is the Education Metaverse? We must first consider what teachers and students can do in the classroom and play the meaning of imaginative and out of nothing. As seen from the seventh layer of the Metaverse, the fourth to seventh layers are built by countries, enterprises, various forms of communities, and engineers to help ordinary people. Teachers and students cannot pass through the classroom and participate in teaching activities. This paper believes that on the scene of the Education circle, the best place to work hard is to imagine and discuss the first to third layers, become participants, sharers, promoters, and creators, and create innovative economic development after infrastructure construction is where Education can exert its irreplaceable influence.

Just like, for large-scale public constructions such as highways, it is unlikely that private forces can afford the maintenance and operation of streets. However, when the infrastructure of roads is built, the general private logistics industry, electronic business industry, and supply chain can flourish, so the economic activities developed by infrastructure construction can be operated by private enterprises.

Therefore, the creator's economic activity is to make teachers think about how to lead students to explore and discover exciting things in life so that they can be mapped into our virtual world and then integrate the concepts of the first and second layers, Creating "experiential communication."

What is "experiential communication"? For example, it is complete if you buy a bowl of braised pork rice and go home and sit in front of the TV to eat. But if, instead, you find a group of good friends and eat a bowl of braised pork rice in a cozy restaurant, you are also complete. Still, it is not only your stomach that is taken care of at that time. If you are delighted during the chat, You have had a lot of conversations and exchanges and even created the possibility of continuing other cooperation in the future. In that dinner, not just eating braised pork rice, you felt physical and mental pleasure, self-cultivation improved, and new values were explored, not just Eat enough.

\begin{figure}[ht]
\includegraphics[width=8.8cm]{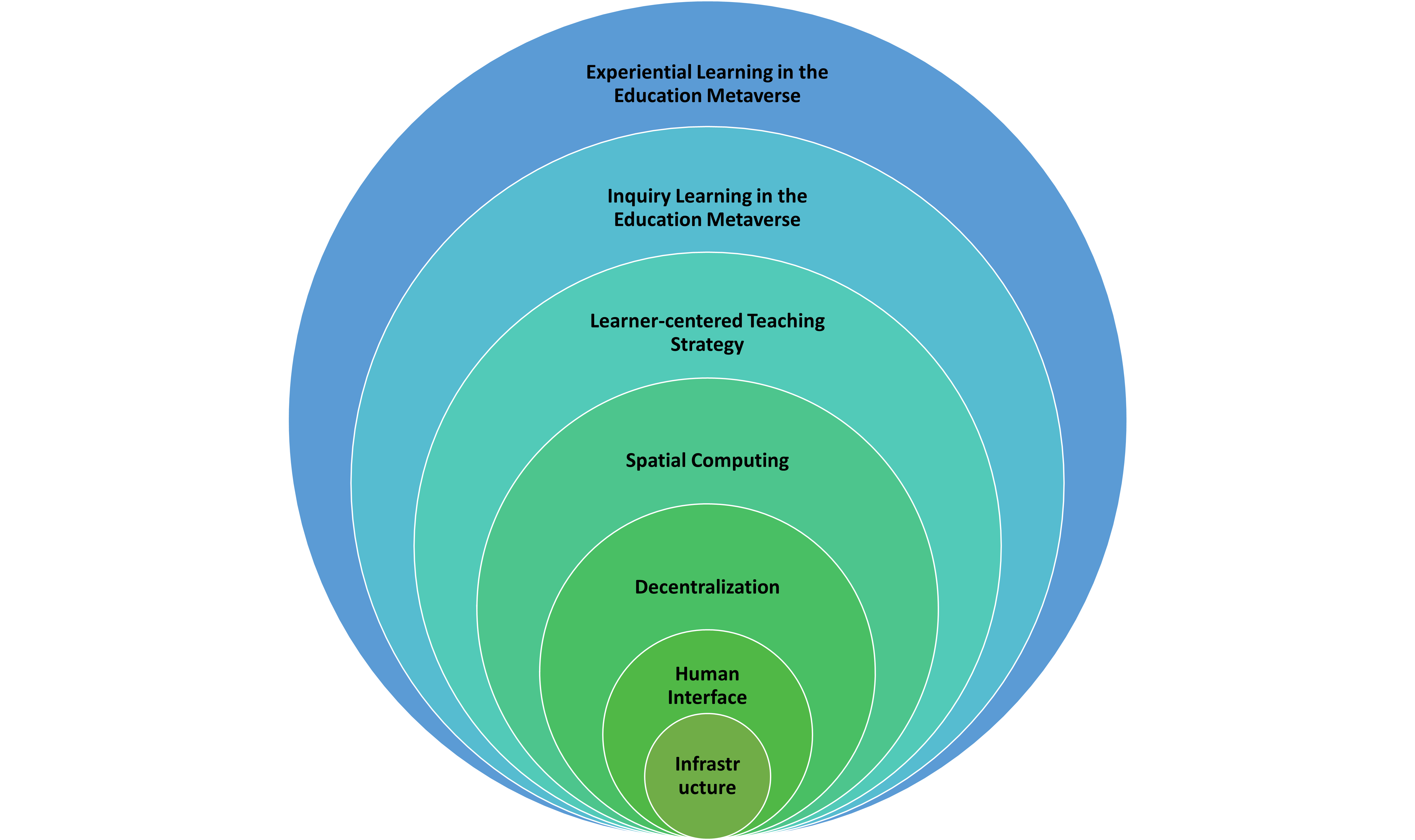}
\caption{The Value Chain of Education Metaverse.}
\label{pecu}
\end{figure}

This is what this paper calls "experiential communication." The following detailed description is carried on the lower four layers of the seventh layer of the Metaverse, and the upper three layers are replaced to become the Education Metaverse. Figure~\ref{pecu} shows the Value Chain of Education Metaverse and the detiles are as follows:
\begin{enumerate}
    \item Experience: Experiential Learning in the Education Metaverse. If we start from the idea of the metaverse value chain defined by Jon Radoff, the imagination of the educational metaverse described in this article is to co-create an exciting and meaningful work or project for teachers and students. The interesting thing is that the world needs to be solved. Teachers and students will try to explore together so that this series of changes can be linked together and become a solution that can improve society and contribute.
    \item Discovery: Inquiry Learning in the Education Metaverse. Because of the co-creation of teachers and students and the participation of learners with unlimited time and space backgrounds, they are given full opportunities to publish, discuss and operate, and create changes in learners' cognitive behavior, and then experience and learn scientific knowledge, scientific attitude and scientific skills, make the application value corresponding to the use of this skill, and continue to share and lead the trend. To achieve the goal of starting from cognition, affection, and abilities to the ultimate education metaverse and to create value.
    \item Creator Economy: Learner-centered Teaching Strategy. To provide learners not only learners but also independent creators, sharers, and promoters, with learners as the main body, teachers should help learners to focus more on creating to gain cognition and affection. In addition to basic abilities such as skills, it also realizes the value of creation. It returns the ownership of creation and learning initiative to learners so that they can directly verify through the blockchain platform on the Internet. What is the value of learning? It helps learners to think about the meaning of learning, find out the intrinsic motivation for learning, and then achieve high-level thinking of continuous learning.
    \item Spatial Computing: In addition to the currently recognized Metaverse using VR, AR, and MR, it echoes the original description of "Experience Layer," which is "But the Metaverse is not 3D or 2D, or even necessarily graphical."
    \item Decentralization: This is one of the most prominent features of blockchain. The ideal Metaverse should be a fairer, more transparent, decentralized, and decentralized world.
    \item Human Interface: This is not limited to using VR, AR, MR. In Education Metaverse, mobile phones, laptops, tablets, 3D floating interactive shadow devices, smart glasses, wearable devices, etc., all belong to A human-machine interface device.
    \item Infrastructure: To realize the characteristics of the Metaverse mentioned above, faster and more popular network infrastructure, more extensive capacity cloud storage services, more powerful image computing equipment, high-endurance durable batteries, micro-perceptrons, and MEMS Systems, etc., all belong to the category of infrastructure.
\end{enumerate}

To realize the Experience: Experiential Learning in the Education Metaverse and the Discovery: Inquiry Learning in the Education Metaverse, the main task is to provide learners not only learners but also independent creators, sharers, and promoters. This paper offers the following subsection to check whether the teaching strategies suit learner needs.

\subsection{Kano Analysis Combined with Likert Scale (KALS)}
Professor Noriaki Kano proposed the KANO Model in 1984. It mainly analyzes customer satisfaction to clarify whether respondents need such designs, functions, and services~\cite{sauerwein1996kano}. This paper presents the Kano analysis combined with the Likert scale (KALS) to understand the students. The KALS is a learner-centered thinking test model that analyzes what learners like to learn. Our KALS is as follows:

\begin{table}[ht]
\resizebox{\columnwidth}{!}{%
\small
\begin{tabular}{|cc|ccccc|}
\hline
\multicolumn{2}{|c|}{\multirow{2}{*}{\begin{tabular}[c]{@{}c@{}}Curriculum Designed \\ to Meet Student Needs\end{tabular}}} &
  \multicolumn{5}{c|}{Feature is absent} \\ \cline{3-7} 
\multicolumn{2}{|c|}{}     & \multicolumn{1}{c|}{1} & \multicolumn{1}{c|}{2} & \multicolumn{1}{c|}{3} & \multicolumn{1}{c|}{4} & 5 \\ \hline
\multicolumn{1}{|c|}{\multirow{5}{*}{\begin{tabular}[c]{@{}c@{}}Feature\\is\\ present\end{tabular}}} &
  1 &
  \multicolumn{1}{c|}{Q} &
  \multicolumn{1}{c|}{R} &
  \multicolumn{1}{c|}{R} &
  \multicolumn{1}{c|}{R} &
  R \\ \cline{2-7} 
\multicolumn{1}{|c|}{} & 2 & \multicolumn{1}{c|}{M} & \multicolumn{1}{c|}{I} & \multicolumn{1}{c|}{I} & \multicolumn{1}{c|}{I} & R \\ \cline{2-7} 
\multicolumn{1}{|c|}{} & 3 & \multicolumn{1}{c|}{M} & \multicolumn{1}{c|}{I} & \multicolumn{1}{c|}{I} & \multicolumn{1}{c|}{I} & R \\ \cline{2-7} 
\multicolumn{1}{|c|}{} & 4 & \multicolumn{1}{c|}{M} & \multicolumn{1}{c|}{I} & \multicolumn{1}{c|}{I} & \multicolumn{1}{c|}{I} & R \\ \cline{2-7} 
\multicolumn{1}{|c|}{} & 5 & \multicolumn{1}{c|}{O} & \multicolumn{1}{c|}{A} & \multicolumn{1}{c|}{A} & \multicolumn{1}{c|}{A} & Q \\ \hline
\end{tabular}%
}
\caption{KALS Evaluation Table. 1: Very Disagree; 5: Very agree.}
\end{table}
Curriculum Designed to Meet Student Needs:
\begin{enumerate}
    \item M: Must-be.
    \item Q: Questionable.
    \item A: Attractive.
    \item I: Indifferent.
    \item R: Reverse.
    \item O: Outstanding.
\end{enumerate}
There are two equations to calculate Satisfaction Index and Dissatisfaction Index (DI).

\begin{align}
    SI &= \frac{A+O}{A+O+M+I},\\
    DI &= \frac{(O+M)\times(-1)}{A+O+M+I}.
\end{align}

\begin{figure}[ht]
\includegraphics[width=8.8cm]{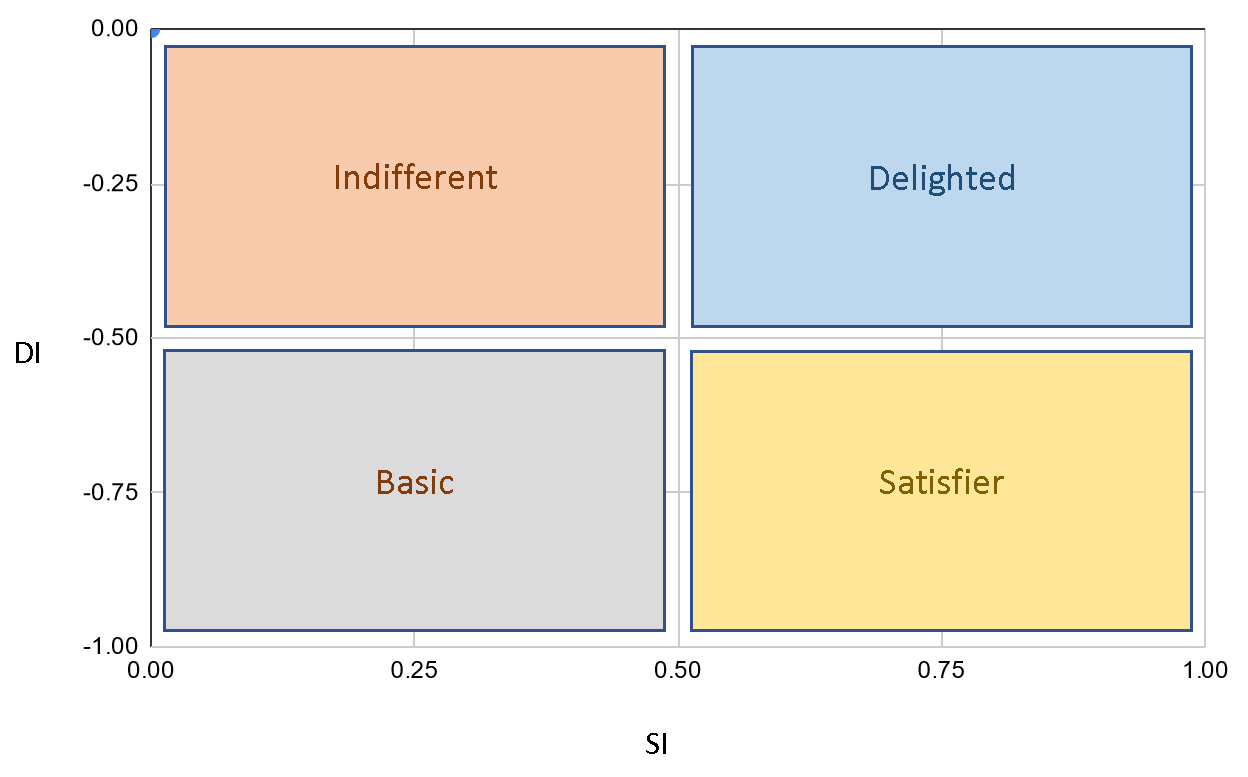}
\caption{The two-dimensional plane scatters plot of the KALS analysis.}
\label{KALS}
\end{figure}

Figure~\ref{KALS} shows that the two-dimensional plane scatters plot of the KALS analysis has four areas to show whether the feature meets student needs.

That's the basics, and now that we know what the KALS model is and where it comes from, let's look at how we can use it to think about Human Based Learning (HBL) for running a Learner-centered teaching strategy, innovation, and what learners like to learn.

We discussed with students the Value Chain of the Education Metaverse. And then, we use the following two questions to ask 46 students from 10 to 18 years old:
\begin{enumerate}
    \item Feature is present: The ability to turn 2D into 3D can help you understand the meaning of the Metaverse.
    \item Feature is absent: No ability to turn 2D into 3D, you also can understand the meaning of the Metaverse.
\end{enumerate}

We get the answers as follows:
\begin{table}[ht]
\resizebox{\columnwidth}{!}{%
\begin{tabular}{|cc|ccccc|}
\hline
\multicolumn{2}{|c|}{} &
  \multicolumn{5}{c|}{Feature is absent} \\ \cline{3-7} 
\multicolumn{2}{|c|}{\multirow{-2}{*}{\begin{tabular}[c]{@{}c@{}}Curriculum Designed \\ to Meet Student Needs\end{tabular}}} &
  \multicolumn{1}{c|}{1} &
  \multicolumn{1}{c|}{2} &
  \multicolumn{1}{c|}{3} &
  \multicolumn{1}{c|}{4} &
  5 \\ \hline
\multicolumn{1}{|c|}{} &
  1 &
  \multicolumn{1}{c|}{\cellcolor[HTML]{C0C0C0}0} &
  \multicolumn{1}{c|}{\cellcolor[HTML]{C0C0C0}0} &
  \multicolumn{1}{c|}{\cellcolor[HTML]{C0C0C0}0} &
  \multicolumn{1}{c|}{\cellcolor[HTML]{C0C0C0}0} &
  \cellcolor[HTML]{C0C0C0}0 \\ \cline{2-7} 
\multicolumn{1}{|c|}{} &
  2 &
  \multicolumn{1}{c|}{\cellcolor[HTML]{C0C0C0}0} &
  \multicolumn{1}{c|}{\cellcolor[HTML]{C0C0C0}0} &
  \multicolumn{1}{c|}{\cellcolor[HTML]{C0C0C0}0} &
  \multicolumn{1}{c|}{\cellcolor[HTML]{C0C0C0}0} &
  \cellcolor[HTML]{C0C0C0}0 \\ \cline{2-7} 
\multicolumn{1}{|c|}{} &
  3 &
  \multicolumn{1}{c|}{\cellcolor[HTML]{C0C0C0}0} &
  \multicolumn{1}{c|}{\cellcolor[HTML]{C0C0C0}0} &
  \multicolumn{1}{c|}{\cellcolor[HTML]{C0C0C0}3} &
  \multicolumn{1}{c|}{\cellcolor[HTML]{C0C0C0}1} &
  \cellcolor[HTML]{C0C0C0}0 \\ \cline{2-7} 
\multicolumn{1}{|c|}{} &
  4 &
  \multicolumn{1}{c|}{\cellcolor[HTML]{C0C0C0}0} &
  \multicolumn{1}{c|}{\cellcolor[HTML]{C0C0C0}5} &
  \multicolumn{1}{c|}{\cellcolor[HTML]{C0C0C0}10} &
  \multicolumn{1}{c|}{\cellcolor[HTML]{C0C0C0}17} &
  \cellcolor[HTML]{C0C0C0}0 \\ \cline{2-7} 
\multicolumn{1}{|c|}{\multirow{-5}{*}{\begin{tabular}[c]{@{}c@{}}Feature \\ is \\ present\end{tabular}}} &
  5 &
  \multicolumn{1}{c|}{\cellcolor[HTML]{C0C0C0}0} &
  \multicolumn{1}{c|}{\cellcolor[HTML]{C0C0C0}1} &
  \multicolumn{1}{c|}{\cellcolor[HTML]{C0C0C0}2} &
  \multicolumn{1}{c|}{\cellcolor[HTML]{C0C0C0}1} &
  \cellcolor[HTML]{C0C0C0}6 \\ \hline
\end{tabular}%
}
\caption{KALS Evaluation Table. The gray area is the answers statistical combin with two questions. 1: Very Disagree; 5: Very agree.}
\end{table}

The SI and DI are:
\begin{align*}
    A &= 1+2+1=4,\\
    O &= 0,\\
    M &= 0,\\
    I &= 5+10+17+3+1=36,\\
    SI &= \frac{A+O}{A+O+M+I}=\frac{4+0}{4+0+0+36}=0.1,\\
    DI &= \frac{(O+M)\times(-1)}{A+O+M+I}=\frac{(0+0)\times(-1)}{4+0+0+36}=0.
\end{align*}

\begin{figure}[ht]
\includegraphics[width=8.8cm]{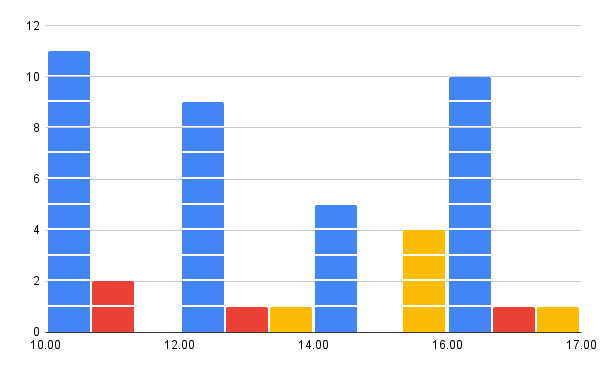}
\caption{The horizontal axis is the age and vertical axis is the counter of the age. The I (Indifferent) is blue box, the A (Attractive) is red box, and the Q (Questionable) is yellow box.}
\label{chart}
\end{figure}

Figure~\ref{chart} shows the answer types of 46 students from 10 to 18 years old. We can see the different ages students can understand the meaning of the Metaverse without using tools to turn 2D into 3D. The case demonstrates how to use the KALS for the curriculum activities of the HBL strategy. In this case, we get that the ability to turn 2D into 3D is indifferent. Suppose the teacher wants to know whether the curriculum activities can return to the needs of the students instead of the teacher judging the students' learning effectiveness from the student's learning evaluation. In that case, a learner-centered learning model is the first step to taking the learner as the main body. The KALS explores the characteristics of the Educational Metaverse, allowing teachers and students to develop more clearly how to conduct in-depth discussions through the features of the education metaverse so that education 4.0 can bring opportunities for realization due to the gradual development of the metaverse, not just the integration of VR, AR, and MR into teaching, but no specific transformation.

In the subsection, we check whether the teaching strategies suit learner needs. This paper provides the three steps of the Learner-centered teaching strategy in the following subsection.

\subsection{The Three Steps of Learner-centered Teaching Strategy}
This paper proposes Human Based Learning (HBL) for running Learner-centered Teaching Strategy. The HBL teaching strategy is into three steps as follows:
\begin{enumerate}
    \item Learner's spontaneous behavior: Most students' unexpected behaviors are defined as learning motivation. How to induce learning motivation is the primary key at this stage. Self-awareness at this stage needs to use the KALS pioneered in this paper. The analysis of the "table" to understand the issues or teaching methods given by students and teachers are classified as unnecessary, necessary, and wanted. The teaching method can also closely relate to the discussion topic and the knowledge content.
    \item Stimulation and interaction with the learning environment: to understand the appropriate operation method for students to carry out the concept of the original universe of the second stage of the virtual world and to be able to try and familiarize themselves with the task-oriented way in the virtual world. After becoming familiar with the interface operation, the masterworks appreciation and extraction elements are carried out. The masterworks appreciation and extraction elements will be extracted in these two trim stages due to the need to find suitable works. The loop method will be repeated to find relevant results and features, then rebate tasks to solve and create.
    \item The learner actively participates in the learning: There are many factors in the learner's participation in the teaching and the learner's spontaneous behavior in the first stage. For active involvement in the education metaverse, it is necessary to check whether the learning motivation is consistent. The interactive report is different from the previous reports. A significant factor is the application of additional equipment in the classroom, whether it is because of the equipment or equipment that students are willing to try to report in this way.
\end{enumerate}

\section{The Features of Education Metaverse}
\subsection{Experiential Learning in the Education Metaverse}
In the field of Education, now it is no longer possible to instill one-way knowledge (just for the sake of being full), but to begin to expect that learning is something that can advance from the consideration of all-round development of body and mind, that is, "experiential communication." If we start from the idea of the Metaverse value chain defined by Jon Radoff, the imagination of the Education Metaverse described in this article is to create an exciting and meaningful work or project for teachers and students. The interesting thing is that the world needs to be solved. Teachers and students work together to try to explore so that this series of changes can be linked together and become a solution that can improve society. Contributed, it is a complete "Education Metaverse of experiential learning."
\begin{enumerate}
    \item Are we in a digital environment where teachers, students, and more participants outside the classroom can co-create?
    \item Is the issue a problem that needs to be solved worldwide?
    \item Can this sequence of operational situations be recorded?
    \item The co-creation content can continue to grow naturally due to the participation of more learners.
    \item It can develop into a phenomenon of social immersion due to the proliferation and interaction of the contents.
\end{enumerate}
\subsection{Inquiry Learning in the Education Metaverse}
Because of the co-creation of teachers and students and the participation of learners with unlimited time and space backgrounds, they are given full opportunities to publish, discuss and operate, and create changes in learners' cognitive behavior, and then experience and learn scientific knowledge, scientific attitude and scientific skills, make the application value corresponding to the use of this skill, and continue to share and lead the trend. To achieve the goal of starting from cognition, affection, and abilities to the ultimate Education Metaverse and to create value.
\begin{enumerate}
    \item Is it possible for participants to adequately publish, discuss and operate?
    \item Can it create a change in the learner's cognitive behavior?
    \item Can learners experience and learn scientific knowledge, attitude, and skills and create application value corresponding to such skills?
    \item Is it possible for learners to identify with their creations and continue to share the trends led by them?
    \item Can it achieve the goal of starting from cognition, affection, and skills to the ultimate Education Metaverse and creating value?
\end{enumerate}
\subsection{Learner-centered Teaching Strategy}
To provide learners not only learners but also independent creators, sharers, and promoters, with learners as the main body, teachers should help learners to focus more on creation to gain cognition and affection. In addition to basic abilities such as skills, it also realizes the value of innovation. It returns the ownership of creation and learning initiative to learners so that they can directly verify through the blockchain platform on the Internet. What is the value of learning? It helps learners to think about the meaning of learning, find out the intrinsic motivation for learning, and then achieve high-level thinking of continuous learning.
\begin{enumerate}
    \item Can the results of the two-dimensional plane scatter plot of the KALS analysis method be?
    \item Can the needs of learners identify from the analysis method of the KALS?
    \item Can it guide learners to become independent creators, sharers, and promoters?
    \item Is it possible to enable learners to focus more on basic abilities such as cognition, affection, and skills and to realize creative value?
    \item Can learners develop learning initiatives by acquiring creative ownership?
\end{enumerate}

\section{Conclusion}
The 2022 World Economic Forum closed on May 26 and launched the "Education 4.0" people-centered recovery initiative. From now on, the teaching model must be learner-centered, make good use of technology and teaching method innovation, and help learners have a more comprehensive range of skills, adaptability, and resilience to face the unknown world in response to the Fourth Industrial Revolution.

The unknown world is driven by a group full of imagination and execution. Starting from the Metaverse value chain defined by Jon Radoff, we can see that the future world can be thought of from the universe's perspective. The universe is a vast conscious body, representing that each small unit in this aware body carries information and signals and can communicate and exchange freely. Meta is the abbreviation of Metadata, which is the data that describes the data. Humans are also a unit in the universe. A universe created from human imagination that can create value through information exchange and communication is the Metaverse, which is the first chapter of the new world that has come.

Because the Education Metaverse is "teacher and students can jointly create any thinking system that can generate value exchange," even if the teacher does not have VR, AR, or MR technology equipment, what kind of presentation method should be used by the participants and the co-creation Students make their own decisions according to local conditions, as long as they can design a teaching plan that conforms to the Education Metaverse spirit so that after entering the society, students can become independent and become an experience of the learning process of all the abilities they have as a human being, to meet the needs of learning Learner-centered teaching strategy to achieve the learner-centered goal of Education 4.0. It is the core value and development goal of the Education Metaverse.

This paper currently focuses on defining the Education Metaverse, proposes relevant measurement indicators, confirms the characteristics of the Education Metaverse, and hopes to introduce broader discussions and reflections. In the future, we will continue to introduce different technological tools to explain how to apply them in the Education Metaverse, including investing in blockchain technology to achieve "digital avatars" and a decentralized "consensus value system," as well as more different Education Metaverse lesson plans. Discuss and guide more teachers to confidently participate in the lesson plan design of the Education Metaverse.

%\bibliographystyle{IEEEtran}      % mathematics and physical sciences
%\bibliography{ref}% common bib file


\begin{IEEEbiography}[{\includegraphics[height=1in,clip,keepaspectratio]{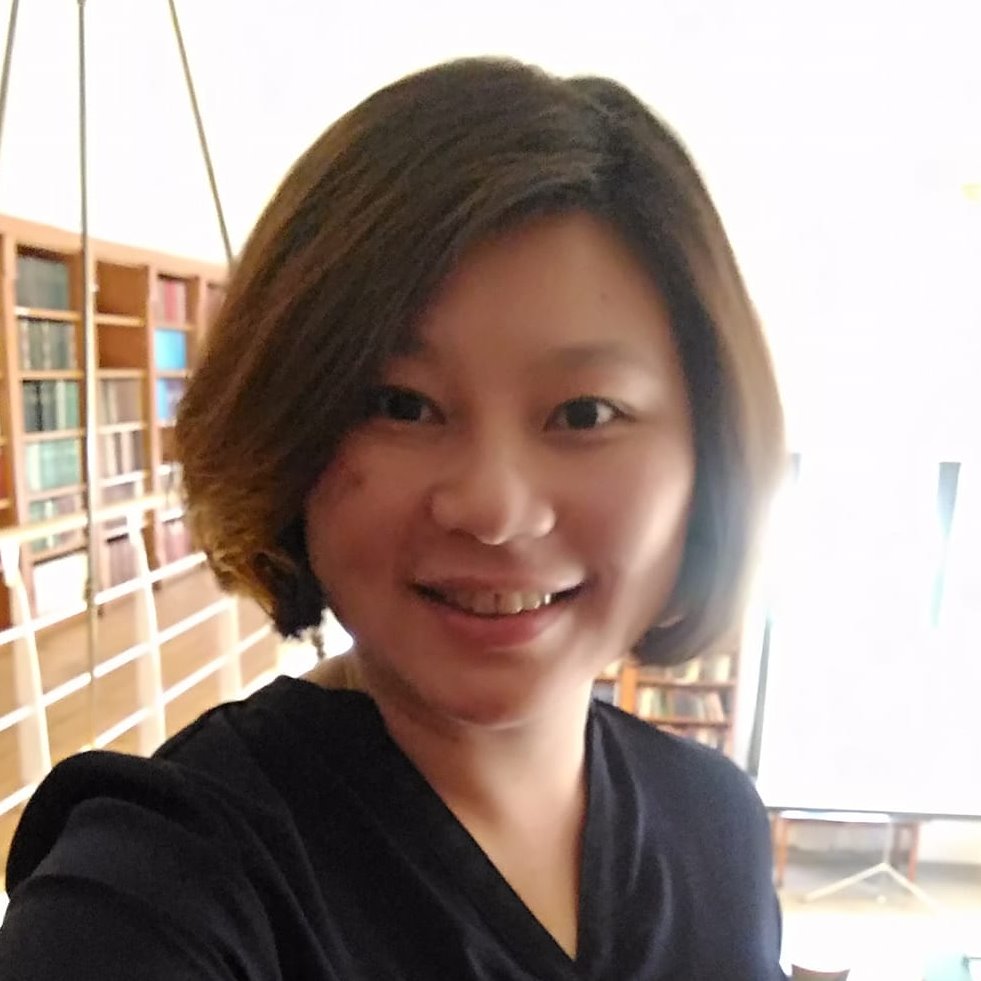}}]{Yun-Cheng Tsai} is an professor at the Department of Technology Application and Human Resource Development at the National Taiwan Normal University. Her faculty experiments include the Department of Data Science at Soochow University and the Center for General Education at National Taiwan University. She served as a consultant at the China Trust Trading Office, a pre-doctoral researcher at the Max Planck Institute for the History of Science in Germany, and a software engineer at Numerix. Specialties are data science, fintech, and blockchain.
\end{IEEEbiography}


%\vfill


\end{document}